\documentclass[aps,prb,twocolumn,superscriptaddress,nofootinbib]{revtex4-2}
\usepackage[english]{babel}
\usepackage{bm}
\usepackage{amsmath}
\usepackage{amssymb}
\usepackage{textcomp}
\usepackage{dutchcal}
\usepackage[colorlinks=true, allcolors=cyan]{hyperref}
\usepackage{diagbox}
\usepackage{multirow}
\usepackage{threeparttable}
\usepackage{xfrac}
\usepackage{xcolor}

\usepackage{ulem}

\begin{document}

\title{Quasi-1D Planar Magnetic Topological Heterostructure}

\author{Z.Z. Alisultanov}
\email{zaur0102@gmail.com}
\affiliation{Abrikosov Center for Theoretical Physics, MIPT, Dolgoprudnyi, Moscow Region 141701, Russia}
\affiliation{L.D. Landau Institute for Theoretical Physics, acad. Semenova av.1-a, Chernogolovka
142432, Russia}
\affiliation{Amirkhanov Institute of Physics of Dagestan Federal Research Centre of RAS, Makhachkala 367003, Russia}

\begin{abstract}
We theoretically introduce a quasi one dimensional magnetic heterostructure of alternating 2D topological and normal insulator strips. Such a structure can be realized by special fabrication techniques or may emerge naturally in twisted bilayers of certain transition metal dichalcogenides. Its low-energy physics is governed by a hybrid Hamiltonian intertwining the Su-Schrieffer-Heeger and Shockley models, with spin-momentum locking and local Zeeman splitting. Symmetry analysis places it in class AIII, characterized by chiral symmetry and a $\mathbb{Z}$ topological invariant. Computing the winding number from the block-off-diagonal structure of the Hamiltonian reveals topological phases characterized by invariants $\nu = 0$, $1$, and $2$. Furthermore, a single magnetic defect acts as a sensitive local probe, whose in-gap spectrum provides a spectroscopic fingerprint to distinguish global topological phases. We show that the frequency-dependent optical absorption induced by such a defect exhibits distinct features in the topological and trivial phases, offering an independent, experimentally accessible local indicator of the global topological order. Extending the platform to a multilayer geometry uncovers a nonsymmorphic projective symmetry that gives rise to Möbius band topology. 
\end{abstract}

\maketitle

\section{Introduction}

The discovery of topological insulators (TIs) has introduced a paradigm where the electronic bulk-boundary correspondence dictates the emergence of robust states protected against disorder and perturbations~\cite{Hasan2010, Qi2011, Moore2010}. In two dimensions, this is exemplified by the quantum spin Hall (QSH) effect, hosting helical edge states immune to backscattering due to time-reversal symmetry (TRS)~\cite{Kane2005, Bernevig2006, Konig2007}. After their experimental realization, these materials have been recognized as promising platforms for observing nontrivial electronic and thermoelectric effects (see,e.g.~\cite{olshanetsky2023observation,Xu,Fu20,alisultanov2025thermoelectric}). A central theme in modern condensed matter physics is the artificial engineering of topological states through heterostructuring~\cite{otrokov2019prediction,deng2020quantum,jiang2020concurrence,liu2023magnetic,wan2024layered,Alisultanov_multilayer,alisultanov2025theory}, where the interplay between topology and geometry gives rise to novel phenomena beyond the intrinsic properties of the constituent materials.

\begin{figure}[t]
    \centering
    \includegraphics[width=1\linewidth]{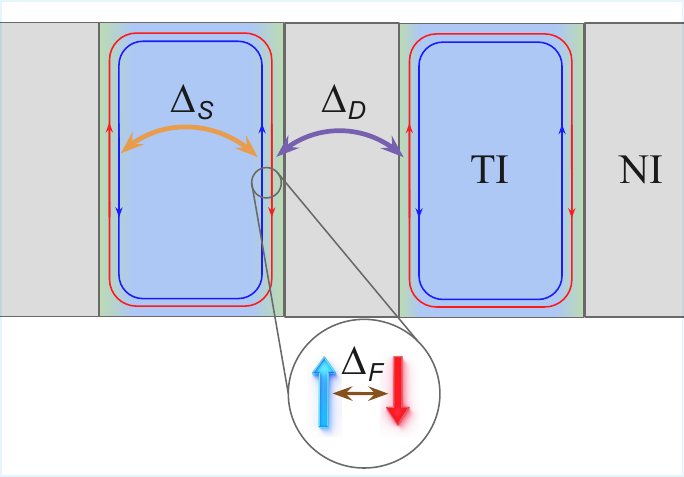}
    \caption{Schematic illustration of the proposed model. The device is a quasi-1D ribbon comprising an alternating sequence of 2D topological insulator (TI) and normal insulator (NI) segments. Magnetic impurities (green) are assumed to be present at each TI-NI interface. The effective coupling between the counter-propagating chiral edge modes is described by the tight-binding Hamiltonian in Eq.~\eqref{Hamiltonian}, with the hopping parameters defined as follows: $\Delta_S$ (tunneling across a TI segment), $\Delta_D$ (tunneling across a NI segment), and $\Delta_F$ (direct on-edge spin-flip tunneling).}
\label{fig:skt}
\end{figure}

A particularly rich direction involves patterning two-dimensional systems into quasi-one-dimensional (quasi-1D) structures to engineer novel topological phases. The relevance of this approach is highlighted by planar Josephson junctions, where such confinement has recently enabled the pursuit of topological superconductivity and Majorana zero modes~\cite{lesser2021phase,pientka2017topological,fornieri2019evidence}. Inspired by this platform, we consider architectures where the chiral edge states of 2D TI segments can hybridize via two distinct tunneling pathways: through the bulk of adjacent TI regions ($\Delta_S$) and across the intervening barriers of 2D normal insulators (NI) ($\Delta_D$) (see~Fig.~\ref{fig:skt}). This inherent dimerization naturally maps onto the Su-Schrieffer-Heeger (SSH) model~\cite{Su1979, Asboth2016}, but is enriched by the essential addition of spin-momentum locking. Furthermore, introducing magnetic impurities at the interfaces~\cite{liu2009magnetic,gao2023quantum,belopolski2025synthesis} breaks TRS and allows for control over spin-dependent transport through local, on-edge spin-flip scattering ($\Delta_F$), potentially driving the system into magnetic topological phases.

Importantly, this alternating stripe geometry can appear naturally in recently discovered systems that exhibit one-dimensional moiré patterns. Examples include graphene nanoribbons on hBN that exhibit periodic domain structures ~\cite{okumura2026one}, twisted bilayer $1T'-\text{WTe}_2$ forming 1D superlattices of alternating stacking domains~\cite{kawakami2026one}, and strain-engineered $\text{MoSe}_2-\text{WSe}_2$ heterobilayers where uniaxial strain transforms triangular moiré patterns into 1D arrays~\cite{zhao2025control}. Given that monolayer $1T'-\text{WTe}_2$ is a quantum spin Hall insulator~\cite{tang2017quantum}, strain or moiré patterning can locally drive regions into a trivial insulating state, creating an intrinsic analogue of our designed heterostructure. A more detailed discussion of the physical realization of our proposed structure in twisted bilayer systems is provided in Appendix~\ref{Appendix A}. Our findings—particularly the predicted phase diagram and the spectroscopic fingerprint of magnetic defects—can be interesting for studies in these emerging 1D moiré platforms. 

The low-energy physics of such quasi-1D heterostructure (Fig.~\ref{fig:skt}) is governed by a hybrid Hamiltonian that intertwines elements of the SSH model (via $\Delta_S$ and $\Delta_D$) and the Shockley model of interface states (via the on-site magnetic term $\Delta_F$)~\cite{Shockley1939,pershoguba2012shockley,mccann2023catalog}, all while retaining the Dirac nature of the original edge states. The competition between these energy scales, along with a Zeeman splitting ($\Delta_z$), is poised to yield a rich phase diagram. However, a comprehensive symmetry classification and a detailed mapping of the topological phases of such a hybrid SSH-Shockley model remain unexplored.

In this work, we theoretically investigate the electronic and topological properties of the above-mentioned heterostructure (Fig.~\ref{fig:skt}). We derive an effective tight-binding Hamiltonian and perform a complete symmetry analysis. We demonstrate that the system, lacking both time-reversal and particle-hole symmetry but possessing a chiral symmetry, belongs to the Altland-Zirnbauer class AIII~\cite{Altland1997,chiu2016classification}. This classification implies a $\mathbb{Z}$ topological invariant in one dimension. We compute the topological phase diagram by evaluating the winding number $\nu$ from the block-off-diagonal structure of the Hamiltonian, revealing topological phases characterized by invariants $\nu = 0$, $1$, and $2$.

Furthermore, we explore the local effect of a single magnetic defect embedded in a non-magnetic host. We show that the spectrum of the bound states induced within the bulk gap acts as a sensitive local probe: in the topological phase, the defect generates four dispersive states with a protected level crossing, while in the trivial phase, it gives rise to only two states, one of which is pinned to the band edge. This stark difference provides a clear spectroscopic fingerprint for distinguishing the global topological order. To substantiate this fingerprint, we also evaluate the optical absorption due to these defect-bound states and demonstrate that it is sharply distinct in the topological and trivial phases, providing a complementary, experimentally measurable local indicator.

Finally, extending this platform to a multilayer geometry reveals a nonsymmorphic projective symmetry that gives rise to Möbius band topology in the electronic structure.

\section{Model and Symmetry Analysis}

We consider the quasi-one-dimensional heterostructure depicted in Fig.~\ref{fig:skt}. Its low-energy physics is described by a tight-binding Hamiltonian featuring two internal degrees of freedom per unit cell:
\begin{itemize}
    \item \textbf{Spin} ($\sigma$), described by Pauli matrices $\sigma_x, \sigma_y, \sigma_z$ acting on the subspace $\{\uparrow, \downarrow\}$.
    \item \textbf{Edge} ($\tau$), described by Pauli matrices $\tau_x, \tau_y, \tau_z$ acting on the subspace $\{A, B\}$ representing the two distinct edges of the TI strip (unit cell of system).
\end{itemize}
The full basis vector for the $n$-th unit cell is thus $\Psi_n = (c_{n,A,\uparrow},\ c_{n,A,\downarrow},\ c_{n,B,\uparrow},\ c_{n,B,\downarrow})^T$. In this basis, the real-space Hamiltonian is given by:
\begin{gather}
H = \sum_n \Psi_n^\dagger \left( v_F k_y \tau_z \sigma_z + \Delta_F \tau_0 \sigma_x + \Delta_S \tau_x \sigma_0 \right) \Psi_n +\nonumber\\
 \sum_n \left[ \Psi_n^\dagger \Delta_z \tau_0 \sigma_z \Psi_n + \left( \Delta_D \Psi_{n+1}^\dagger \tau_+ \sigma_0 \Psi_n + \text{h.c.} \right) \right],
\label{Hamiltonian}
\end{gather}
where $\tau_\pm=1/2\left(\tau_x\pm i\tau_y\right)$ and the terms represent: $v_F k_y \tau_z \sigma_z$ is the spin-orbit coupled propagation along the edge ($y$-direction), $\Delta_F \tau_0 \sigma_x$ is the spin-flip hybridization within a unit cell, $\Delta_S \tau_x \sigma_0$ is tunneling between edges across a TI segment (intra-cell hopping), $\Delta_D \tau_x \sigma_0$ is tunneling between edges across a NI segment (inter-cell hopping), $\Delta_z \tau_0 \sigma_z$ is Zeeman splitting induced by magnetic impurities. A detailed discussion of tunneling between chiral edge modes in a two-dimensional topological insulator of finite width is provided in Appendix~\ref{Appendix B}. The term $\Delta_F \tau_0 \sigma_x$ describes hybridization between modes on the same edge, arising from local spin-flip scattering off magnetic impurities. While this term is not essential for the main topological features—the same phase diagram can be obtained with $\Delta_F=0$ and $\Delta_z\neq0$—it is retained for completeness of our consideration. For brevity, the tensor product symbol $\otimes$ is omitted throughout. 

The corresponding Bloch Hamiltonian in momentum space is:
\begin{gather}
\mathcal{H}(\mathbf{k}) = v_F k_y \tau_z \sigma_z + \Delta_F \tau_0 \sigma_x + \Delta_z \tau_0 \sigma_z +\nonumber\\
+ \left( \Delta_S + \Delta_D \cos k_x \right) \tau_x \sigma_0+\Delta_D \sin k_x \tau_y \sigma_0.
\label{Hamiltonian momentum}
\end{gather}

This Hamiltonian coincides with the Hamiltonian of the SSH model at $\Delta_F,\Delta_z,k_y=0$. However, unlike the simple SSH model, there are two orbitals at each site. The term $v_F k_y \tau_z \sigma_z$ represents spin-momentum locking along the edges, coupling the spin and orbital degrees of freedom. And at $\Delta_F\neq0$, this Hamiltonian partially includes the Shockley model. Indeed, in the Shockley model each unit cell contains two orbitals with on-site hybridization, and the topological order of the system is governed by the competition between this hybridization and inter-site tunneling~\cite{pershoguba2012shockley,maldonado2024comparing}. Therefore, we can call this Hamiltonian the hybrid SSH-Shockley model.

Alternatively, the model can be viewed as a four-band ladder consisting of two coupled Rice-Mele chains, corresponding to the spin-up and spin-down sectors. The coupling between the chains is provided by the \(\Delta_F\) term, while \(\Delta_z\) acts as an energy offset between the two sectors. This perspective connects our model to previous studies of coupled SSH chains and two-dimensional SSH-like lattices~\cite{liu2017novel,ma2022electronic,agrawal2023cataloging,matveeva2023one}. The presence of chiral symmetry in our model, despite the interchain coupling, makes it an interesting counterexample in this class of systems.

The symmetry properties of the Hamiltonian $\mathcal{H}$ are fundamental to classifying its potential topological phases. We analyze its behavior under time-reversal, particle-hole, chiral, and mirror symmetries. The results are summarized in Table~\ref{table:symmetries}.

\begin{table}[t!]
\centering
\caption{Symmetry classification of the Hamiltonian~\eqref{Hamiltonian momentum}.}
\label{table:symmetries}
\begin{tabular}{lc}
\hline
\textbf{Symmetry} & \textbf{Present} \\
\hline
Time-Reversal ($\mathcal{T}$) & No \\
Particle-Hole ($\mathcal{P}$) & No \\
Chiral ($\mathcal{C}=\tau_z\sigma_y$) & Yes \\
Mirror ($M_x$) & No \\
Mirror ($M_y$) & No \\
\hline
\end{tabular}
\end{table}

\paragraph{Time-Reversal Symmetry (TRS)} is an antiunitary operation represented by $\mathcal{T} = U_T \mathcal{K}$, where $\mathcal{K}$ denotes complex conjugation. For spin-1/2 particles, $U_T = i\tau_0\sigma_y$. The Hamiltonian respects TRS if $\mathcal{T} \mathcal{H}(k_x, k_y) \mathcal{T}^{-1} = \mathcal{H}(-k_x, -k_y)$. The spin-orbit coupling term $v_F k_y \tau_z \sigma_z$ is TRS-invariant. However, both the spin-hybridization term $\Delta_F \tau_0 \sigma_x$ and the Zeeman term $\Delta_z \tau_0 \sigma_z$ change sign under TRS: $\mathcal{T} (\Delta_F \tau_0 \sigma_x) \mathcal{T}^{-1} = -\Delta_F \tau_0 \sigma_x$ and $\mathcal{T} (\Delta_z \tau_0 \sigma_z) \mathcal{T}^{-1} = -\Delta_z \tau_0 \sigma_z$, explicitly breaking TRS.

\paragraph{Particle-Hole Symmetry (PHS)}. A Hamiltonian respects PHS if there exists an antiunitary operator $\mathcal{P}$ such that $\mathcal{P} \mathcal{H}(k_x, k_y) \mathcal{P}^{-1} = -\mathcal{H}(-k_x, -k_y)$. We examine both classes of particle-hole symmetry. As an example, consider the candidate $\mathcal{P}_+ = \tau_y \sigma_y \mathcal{K}$ for $\mathcal{P}_+$ symmetry ($\mathcal{P}^2_+ = 1$). Applying it to the spin-orbit term yields $\mathcal{P}_+ (v_F k_y \tau_z \sigma_z) \mathcal{P}^{-1}_+= -v_F k_y \tau_z \sigma_z$, while the PHS condition requires $+v_F k_y \tau_z \sigma_z$ for this term. Furthermore, the Zeeman term $\Delta_z \tau_0 \sigma_z$ remains invariant under this transformation instead of acquiring a minus sign. A comprehensive analysis shows that no operator $\mathcal{P}_-$ satisfying the PHS condition exists for this Hamiltonian. Thus, the system possesses neither $\mathcal{P}_+$ nor $\mathcal{P}_-$ symmetry.

\paragraph{Chiral Symmetry (CS)}. A Hamiltonian possesses CS if there exists a unitary operator $\mathcal{C}$ such that $\mathcal{C} \mathcal{H}(k_x, k_y) \mathcal{C}^{-1} = -\mathcal{H}(k_x, k_y)$ and $\mathcal{C}^2=1$. We find that the operator $\mathcal{C} = \tau_z \sigma_y$ fulfills this condition. It anticommutes with all terms in the Hamiltonian: $\tau_z \sigma_y$ anticommutes with $\tau_z \sigma_z$, $\tau_0 \sigma_x$, $\tau_0 \sigma_z$, $\tau_x \sigma_0$, and $\tau_y \sigma_0$. 

\paragraph{Mirror Symmetries}. We consider mirror reflections $M_x$ (inversion of $x \rightarrow -x$) and $M_y$ (inversion of $y \rightarrow -y$). The operator $M_x$ is represented by $i \tau_x\sigma_x$ and $M_y$ by $i \tau_0\sigma_y$. The spin-orbit term $v_F k_y \tau_z \sigma_z$ is invariant under both $M_x$ and $M_y$. But the Zeeman term $\Delta_z \tau_0 \sigma_z$ changes sign under both $M_x$ and $M_y$. Consequently, neither $M_x$ nor $M_y$ is a symmetry of the full Hamiltonian~\eqref{Hamiltonian momentum}. As we discuss below, the breaking of these mirror symmetries is crucial for the photocurrent generation in this system.

According to the Altland-Zirnbauer classification \cite{Altland1997}, the Hamiltonian, lacking both TRS and PHS but possessing a chiral symmetry, belongs to \textbf{class AIII}. In one dimension, this class is characterized by a $\mathbb{Z}$ topological invariant. 

At this point, we note a subtlety concerning the dimensionality of the system. The Hamiltonian in Eq. (2) is two-dimensional, but the topological invariant is computed by treating \(k_y\) as a fixed parameter - a standard dimensional reduction procedure \cite{ryu2010topological}. At \(k_y = 0\) and \(\Delta_F = \Delta_z = 0\), an emergent time-reversal symmetry appears, placing the system in class BDI. For generic \(k_y\) and finite magnetic terms, the system belongs to class AIII, as discussed above.

\begin{figure}[t]
    \centering
    \includegraphics[width=\linewidth]{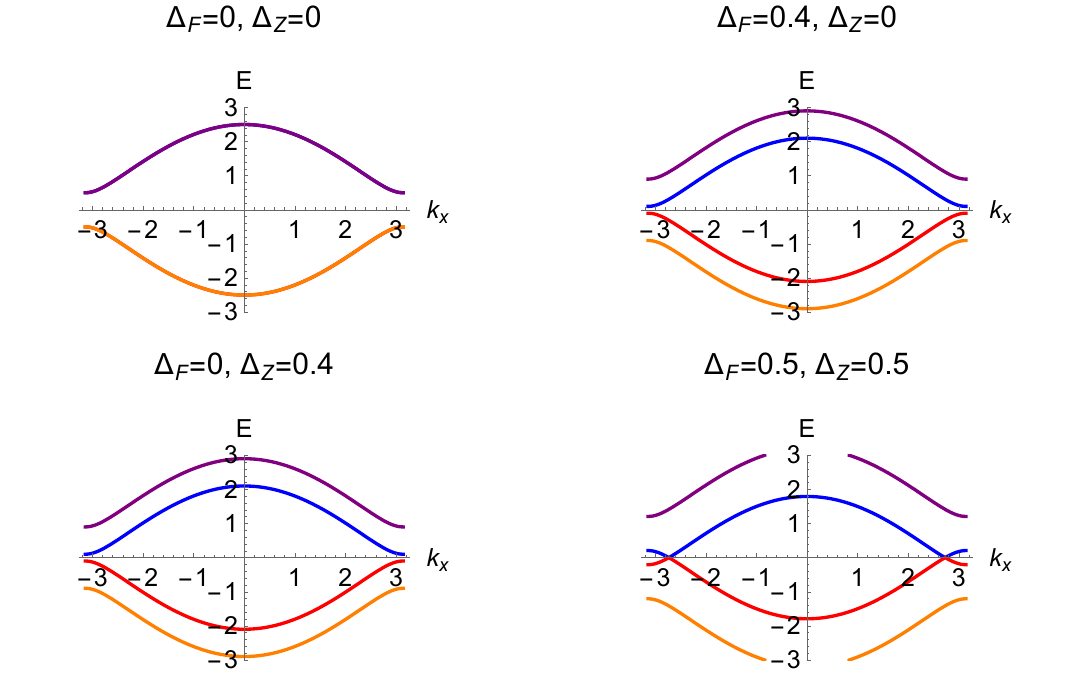}
    \caption{Electronic spectrum~\eqref{eq:spectrum} calculated for various Hamiltonian parameters. Note that the magnetic terms $\Delta_F$ and $\Delta_Z$ produce equivalent band splittings. We have used the following parameter values: $\Delta_S = 1$, $\Delta_D = 1.5$, and $v_F = 1$. The results in these figures are shown for $k_y = 0$. The $k_y$-dependence is discussed in Section IV and presented in Fig.~\ref{new states density of states}.}
    \label{spectrum}
\end{figure}

Beyond the standard Altland-Zirnbauer classification, the Hamiltonian exhibits a unitary symmetry that imposes a fundamental constraint on its spectral structure. We identify the unitary operator $\tau_x$, which satisfies the relation: $\tau_x\mathcal{H}\left(\mathbf{k}\right)\tau_x=-\mathcal{H}\left(-\mathbf{k}\right)$. This is distinct from particle-hole symmetry as the operator is unitary rather than antiunitary. The immediate consequence: for every eigenstate at wavevector $\mathbf{k}$ with energy $E(\mathbf{k})$, there exists a corresponding eigenstate at wavevector $-\mathbf{k}$ with energy $-E(\mathbf{k})$. While the system's topological classification remains that of class AIII due to the presence of chiral symmetry and the absence of antiunitary symmetries, this additional unitary symmetry can have profound implications for the structure of the topological boundary states and the protection of certain degeneracies.

Finally, we note that for $\Delta_F=\Delta_z=0$ and $\Delta_S=\Delta_D$, a hidden (projective) symmetry is restored in our system, defined by the operator $\mathrm{L}_x=L_x\otimes \mathrm{G}=\begin{pmatrix}
0&1\\
e^{ik_x}&0
\end{pmatrix}\otimes\sigma_x$ (pure $L_x$ is not a symmetry). This symmetry corresponds to a translation $L_x$ along the X-axis by half the period of the unit cell combined with a spin flip. As is straightforward to verify, $\left[\mathrm{L}_x,\mathcal{H}\right]=0$ holds in this case. This symmetry with TRS enforce the two-fold degeneracy of bands and appearance of the bands crossing point - a Dirac point - in the Hamiltonian (see, e.g.,~\cite{zhao2020z}).

Next, we diagonalize the Bloch Hamiltonian $\mathcal{H}(k_x, k_y)$. The energy spectrum $E(\mathbf{k})$ is obtained by solving $\det\left( \mathcal{H}(k_x, k_y) - E \mathbb{I}_{4\times4} \right) = 0$:

\begin{widetext}  
\begin{equation}
\label{eq:spectrum}
\begin{aligned}
E_{\pm}^{\pm}\left(k_x,k_y\right)=\pm\sqrt{\Delta^2\left(k_x\right)+\Delta _F^2+v_F^2 k_y^2+\Delta _z^2\pm2 \sqrt{\left(\Delta _F^2+\Delta _z^2\right) \Delta^2\left(k_x\right)+v_F^2 k_y^2 \Delta _z^2}}
\end{aligned}
\end{equation}
\end{widetext}
where $\Delta^2\left(k_x\right)=\Delta _D^2+\Delta _S^2+2 \Delta _D \Delta _S \cos k_x$. Fig.~\ref{spectrum} shows this spectrum. In general, it consists of four non-degenerate bands, a consequence of the broken time-reversal symmetry which lifts all Kramers degeneracies. The terms $\Delta_z \tau_0 \sigma_z,~\Delta_F \tau_0 \sigma_x$ explicitly break time-reversal symmetry, lifting the spin degeneracy and resulting in four distinct bands.

At $k_y=0$ we have the simple expression for the four energy bands:
\begin{equation}
E_{\pm}^{\pm}(k_x)=\pm\bigl|\,R\pm |t(k_x)|\,\bigr|,
\label{eq:spectrum_ky0}
\end{equation}
where $R=\sqrt{\Delta_F^2+\Delta_z^2}$ and $t(k_x)=\Delta_S+\Delta_D e^{-ik_x}$.
The two inner bands, $E_{2}(k_x)=-|R-|t(k_x)||$ and $E_{3}(k_x)=|R-|t(k_x)||$, determine the direct band gap
\begin{equation}
\Delta_{\text{gap}}(k_x)=E_{3}(k_x)-E_{2}(k_x)=2\bigl|\,R-|t(k_x)|\,\bigr|.
\label{eq:direct_gap}
\end{equation}
The \emph{global} bulk gap is the minimum of \eqref{eq:direct_gap} over the Brillouin zone.  
Since $|t(k_x)|$ varies between
\begin{equation}
|t|_{\min}= \bigl|\,|\Delta_S|-|\Delta_D|\,\bigr|, \qquad 
|t|_{\max}= |\Delta_S|+|\Delta_D|,
\label{eq:t_min_max}
\end{equation}
three distinct regimes emerge:

i) \textbf{$R<|t|_{\min}$}: then $R-|t(k_x)|<0$ for all $k_x$ and 
$\Delta_{\text{gap}}=2(|t|_{\min}-R)>0$; the system is gapped.

ii) \textbf{$|t|_{\min}<R<|t|_{\max}$}: there exist two momenta $k_x^*$ satisfying $|t(k_x^*)|=R$.  
At these points $\Delta_{\text{gap}}(k_x^*)=0$ and the bulk gap closes; the system is gapless.

iii) \textbf{$R>|t|_{\max}$}: then $R-|t(k_x)|>0$ for all $k_x$ and 
$\Delta_{\text{gap}}=2(R-|t|_{\max})>0$; the system is again gapped.

These three regimes correspond precisely to the topological phases considered in the next section.

\section{Topological Properties} 

Here we study topological properties of our systems. The Hamiltonian \eqref{Hamiltonian momentum} describes a generalized Su-Schrieffer-Heeger (SSH) model enriched by spin-dependent terms (SSH+Shockley ). The conserved chiral symmetry ensures that the spectrum is symmetric around $E=0$, allowing for the existence of topologically protected zero-energy edge states in a finite open system.

A key observation is that $\mathcal{H}(\mathbf{k})$, possessing chiral symmetry $\mathcal{C} = \tau_z \sigma_y$ with $\mathcal{C}^2 = I$ and $\mathcal{C} \mathcal{H}(\mathbf{k}) \mathcal{C}^{-1} = -\mathcal{H}(\mathbf{k})$, can be brought to a block-off-diagonal form by a unitary transformation. The appropriate unitary matrix is given by:
\begin{equation}
U = \frac{1}{\sqrt{2}}
\begin{pmatrix}
1 & 0 & 1 & 0 \\
i & 0 & -i & 0 \\
0 & 1 & 0 & 1 \\
0 & i & 0 & -i
\end{pmatrix}.
\end{equation}

Applying this transformation, $\mathcal{H}{\text{eff}}(\mathbf{k}) = U^\dagger \mathcal{H}(\mathbf{k}) U$, and subsequently reordering the basis to group states by their chiral eigenvalues, yields the block-off-diagonal form:
\begin{equation}
\mathcal{H}_{\text{eff}}(\mathbf{k}) =
\begin{pmatrix}
0 & Q(\mathbf{k}) \\
Q^\dagger(\mathbf{k}) & 0
\end{pmatrix},
\end{equation}
where $Q(\mathbf{k})$ is a $2 \times 2$ matrix encapsulating the inter-block coupling. This structure is mandated by the chiral symmetry and enables the computation of the topological invariant via the winding number of $Q(\mathbf{k})$. The explicit form of the off-diagonal block $Q(\mathbf{k})$ is found to be:
\begin{equation}
Q(\mathbf{k}) = \begin{pmatrix}
-i\Delta_F + v_F k_y + \Delta_z & \Delta_S + \Delta_D e^{-i k_x} \\
\Delta_S + \Delta_D e^{-i k_x} & i\Delta_F - v_F k_y + \Delta_z
\end{pmatrix}.
\label{eq:Q_matrix}
\end{equation}

For one-dimensional systems in class AIII, the topological invariant is the winding number. In our case we have quasi-one-dimensional system and so
\begin{equation}
\nu(k_y)=\frac{1}{2\pi i}\int_{-\pi}^{\pi} dk_x\,
\operatorname{Tr}\!\left[Q^{-1}(k_x,k_y)\,\frac{\partial Q(k_x,k_y)}{\partial k_x}\right],
\label{eq:winding_def}
\end{equation}
which takes integer values $\nu\in\mathbb{Z}$. 
It can be conveniently expressed through the phase evolution of the determinant:
\begin{equation}
\nu(k_y)=\frac{1}{2\pi}\,\Delta_{k_x}\arg\det Q(k_x,k_y),
\label{eq:winding_det}
\end{equation}
where $\Delta_{k_x}$ denotes the change over the Brillouin zone.

To uncover the phase diagram we first set $k_y=0$; the dependence on $k_y$ will be discussed later. 
With $R^2\equiv\Delta_z^2+\Delta_F^2$ and $t(k_x)=\Delta_S+\Delta_D e^{-ik_x}$, the determinant reads
\begin{equation}
\det Q(k_x)=R^2-t^2(k_x)=-\Delta_D^2\bigl(w-w_+\bigr)\bigl(w-w_-\bigr),
\label{eq:det_factorized}
\end{equation}
where $w=e^{-ik_x}$ and
\begin{equation}
w_{\pm}=-\frac{\Delta_S}{\Delta_D}\pm\frac{R}{\Delta_D}.
\label{eq:w_pm}
\end{equation}
From Eq.~(\ref{eq:winding_det}) the winding number equals the number of roots $w_{\pm}$ that lie inside the unit circle $|w|<1$:
\begin{equation}
\nu=n_++n_-,\qquad n_{\pm}=1\;\; \text{if}\;\; |w_{\pm}|<1,\;\; \text{otherwise}\;\;0.
\label{eq:nu_via_roots}
\end{equation}
The conditions $|w_{\pm}|<1$ translate into
\begin{equation}
|\Delta_S\mp R|<\Delta_D .
\label{eq:conditions}
\end{equation}

\begin{figure}[t]
\centering
\includegraphics[width=\linewidth]{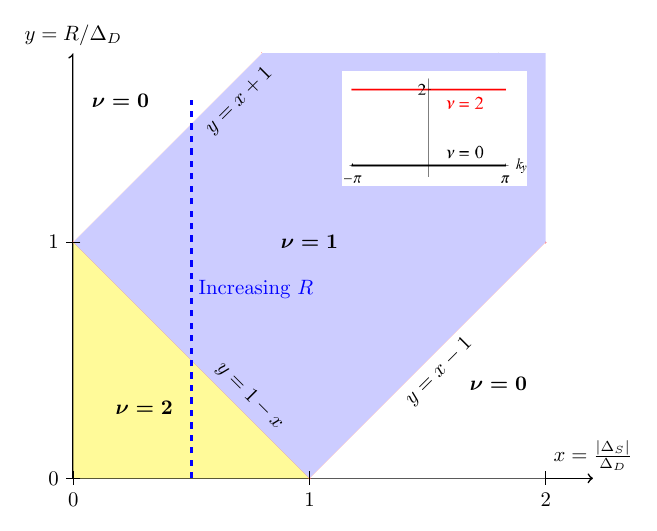}
\caption{
Topological phase diagram of the hybrid SSH--Shockley model at $k_y=0$.
The dimensionless parameters are $x=|\Delta_S|/\Delta_D$ and $y=R/\Delta_D$ with $R=\sqrt{\Delta_F^2+\Delta_z^2}$.
Phases are characterized by the winding number $\nu$.
The dashed line indicates a cut at fixed $x=0.5$, along which increasing the magnetic parameters ($y$) drives the transitions $\nu=2\to1\to0$.
The dot marks the triple point $(x,y)=(1,0)$. Inset: Winding number \(\nu\) as a function of \(k_y\). The red and black lines correspond to the topological (\(\nu = 1\)) and trivial (\(\nu = 0\)) phases, respectively. As shown analytically in Sec. III, \(\nu(k_y)\) remains constant for all \(k_y\) within the gapped phase.
}
\label{fig:phase_diagram}
\end{figure}

Introducing the dimensionless parameters
\begin{equation}
x=\frac{|\Delta_S|}{\Delta_D},\qquad y=\frac{R}{\Delta_D}\;( \ge 0),
\end{equation}
we obtain three distinct topological phases shown in Fig.~\ref{fig:phase_diagram}.

We now comment on the dependence of the topological phase on the transverse momentum \(k_y\). In the gapped phase, we have \(\Delta^2(k_x) > R^2\) for all \(k_x\), where \(R^2 = \Delta_F^2 + \Delta_z^2\). The gap-closing condition at finite \(k_y\) reads
\[
(\Delta^2 - R^2)^2 + v_F^4 k_y^4 + 2 v_F^2 k_y^2 (\Delta^2 + \Delta_F^2 - \Delta_z^2) = 0.
\]
Since \(\Delta^2 > \Delta_F^2 + \Delta_z^2\), we have \(\Delta^2 + \Delta_F^2 - \Delta_z^2 > 0\). Hence all terms in the equation are strictly positive, and no solution exists for any \(k_y\). Therefore, the bulk gap remains open for all \(k_y\), and the winding number \(\nu\) is independent of \(k_y\) as long as the system is in the gapped phase (see Inset on Fig.~\ref{fig:phase_diagram}). This justifies the phase diagram presented in Fig.~\ref{fig:phase_diagram}, which was obtained at \(k_y = 0\).

\textbf{Phase $\nu=2$} (yellow triangle) occurs when both roots lie inside the unit circle:
  \begin{equation}
  y<1-x\quad\text{and}\quad x<1,
  \end{equation}
  i.e. $\Delta_D>|\Delta_S|$ and $R<\Delta_D-|\Delta_S|$.
  In this regime the magnetic terms are weak, and the system effectively consists of two uncoupled SSH copies, each contributing $\nu=\pm1$.

\textbf{Phase $\nu=1$} (blue strip) appears when only one root is inside:
  \begin{equation}
  |x-1|<y<x+1 .
  \end{equation}
  This corresponds to $\Delta_D-|\Delta_S|<R<\Delta_D+|\Delta_S|$.
  Here the magnetic coupling hybridizes the two channels, reducing the total winding number by one. This corresponds to gapless phase with Dirac/Weyl points.

\textbf{Phase $\nu=0$} (white regions) is realized when both roots are outside the unit circle, which happens either for
  \begin{equation}
  y>x+1\quad\text{or}\quad (x>1\;\text{and}\;y<x-1).
  \end{equation}
  The former condition describes strong magnetic terms ($R>\Delta_D+|\Delta_S|$) that completely destroy the topology; the latter corresponds to the trivial SSH limit $|\Delta_S|>\Delta_D$ with negligible $R$.

The phase boundaries are the lines $y=|x-1|$ and $y=x+1$, which meet at the triple point $(x,y)=(1,0)$. 
A direct transition $\nu=2\leftrightarrow0$ occurs only along the line $x=0$ ($\Delta_S=0$), where the intermediate phase $\nu=1$ is absent.

The phase diagram shows that magnetic fields can induce topological transitions.
For a fixed ratio $x<1$, increasing the magnetic parameters (i.e. increasing $y$) moves the system from the $\nu=2$ phase through the $\nu=1$ phase into the trivial $\nu=0$ phase. 
Conversely, for $x>1$ a sufficiently large magnetic term can drive the system from the trivial phase into the topological $\nu=1$ phase before it becomes trivial again at even larger fields.

\begin{figure}
    \centering
    \includegraphics[width=\linewidth]{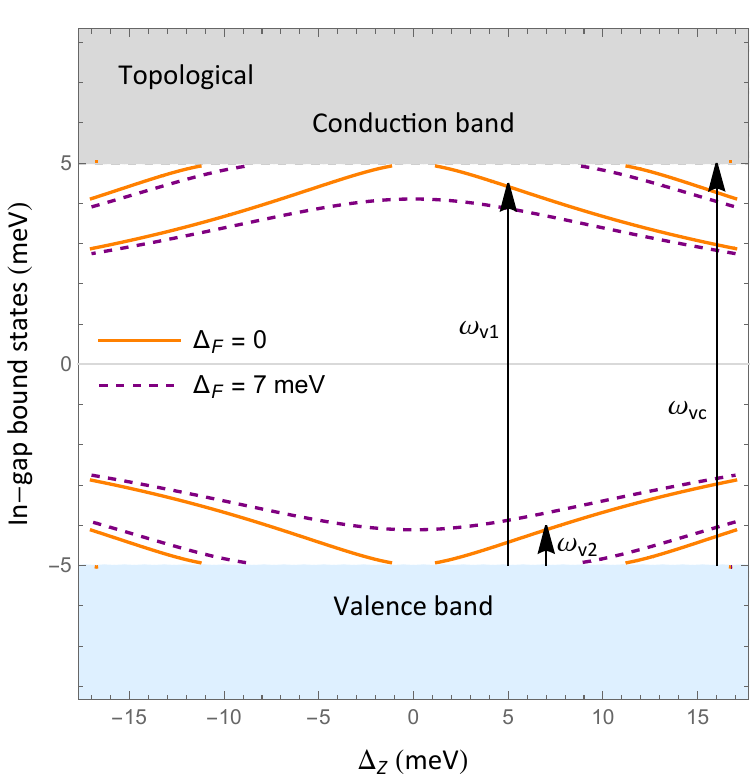}
    \includegraphics[width=\linewidth]{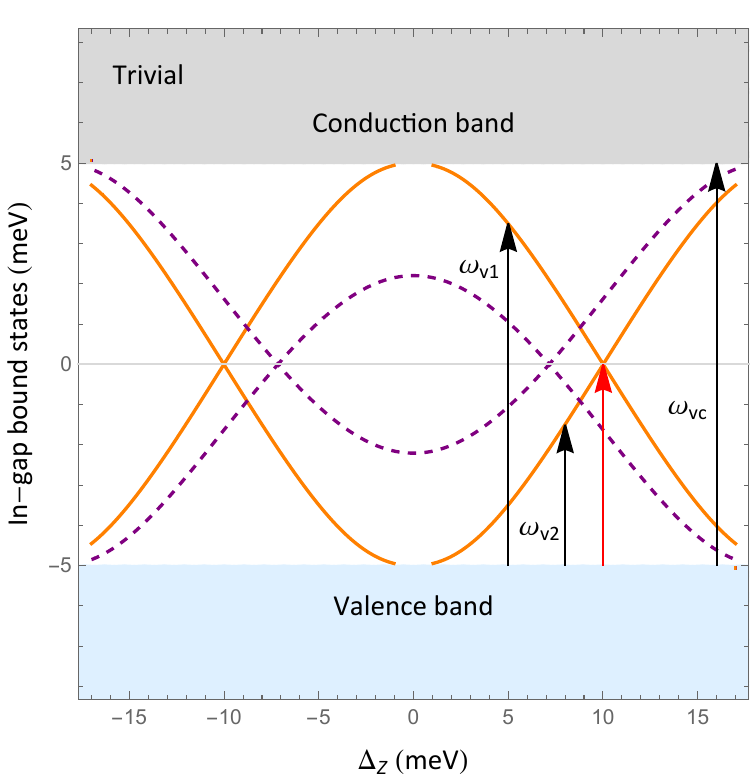}
    \caption{(Color online) Energy spectra of in-gap bound states induced by a single magnetic defect, calculated at $k_y = 0$ as a function of the Zeeman splitting strength $\Delta_Z$ and intra-edge intermodes tunneling magnitude $\Delta_F$. (a) Topological phase ($\Delta_S < \Delta_D$). The introduction of the defect gives rise to four in-gap states. The initially degenerate levels split, and all states exhibit weak dependence on $\Delta_Z$. In contrast to the trivial phase, the bound states in the topological phase don't cross the zero energy. (b) Trivial phase ($\Delta_S > \Delta_F$). Only two in-gap states emerge. While the degeneracy is also lifted, one branch remains independent of $\Delta_Z$ and is pinned precisely at the edge of the bulk band gap. Notably, the in-gap states cross zero energy at a critical value of $\Delta_Z$. The spectra are obtained from the analytical expression in Eq. (\ref{ingap states}).}
    \label{new states into gap}
\end{figure}

\begin{figure}
    \centering
    \includegraphics[width=\linewidth]{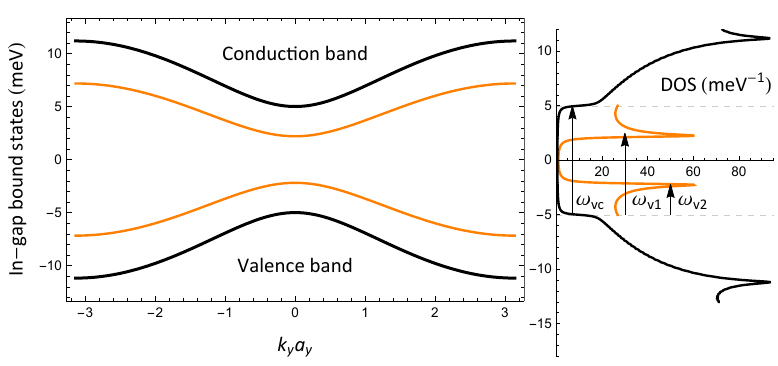}
    \caption{Left panel: $k_y$-dependence of the spectrum of the pristine system (conduction and valence bands - black lines) together with the in-gap defect state for $\Delta_z = 7$ meV (orange lines) in the trivial phase ($\Delta_S = 10$ meV, $\Delta_D = 5$ meV). Right panel: density of states of the conduction, valence, and defect bands. The arrows indicate possible optical transitions from the valence band.}
    \label{new states density of states}
\end{figure}

\section{Single Magnetic TI in non-magnetic structure}

In this section, we investigate the electronic properties of the heterostructure comprising alternating two-dimensional topological and trivial insulators, where only a single topological layer is magnetically doped—effectively creating a magnetic defect in an otherwise nonmagnetic system. Unlike the previously studied fully magnetic configurations that exhibit broken time-reversal symmetry globally, this architecture allows us to probe the local effects of a symmetry-breaking perturbation embedded in a trivial environment. We focus on the emergence and behavior of bound states near the magnetic domain, analyzing their energy dependence on interlayer tunneling couplings in both topological and trivial phases. The hybridization of edge states and the formation of localized modes are examined to uncover the interplay between topology and controlled symmetry breaking.

Recently, the use of defects as probes of topological order has been reviewed in Ref.~\cite{mains2026ordinary}. In contrast to previous work that focused on edge or vortex defects, we consider a single magnetic impurity inside a TI segment and demonstrate that its in-gap spectrum and optical response provide a clear fingerprint of the bulk topology.

We consider a non-magnetic topological insulator system with parameters:
\begin{equation}
\Delta_F = 0, \quad \Delta_z = 0, \quad \Delta_S, \Delta_D \neq 0
\end{equation}
such that the system is in the topological phase ($\nu = 1$).

A single magnetic defect is introduced at site $i_0$ with:
\begin{equation}
\Delta_F(i) = \Delta_F^0 \delta_{i,i_0}, \quad \Delta_z(i) = \Delta_z^0 \delta_{i,i_0}
\end{equation}
where $\delta_{i,i_0}$ is the Kronecker delta function.

The full Hamiltonian becomes:
\begin{equation}
\mathcal{H} = \mathcal{H}_{\text{bulk}} + \mathcal{H}_{\text{defect}}
\end{equation}
with
\begin{gather}
\mathcal{H}_{\text{bulk}} = \sum_n \Psi_n^\dagger \left( v_F k_y \tau_z \sigma_z + \Delta_S \tau_x \sigma_0 \right) \Psi_n +\nonumber\\
+ \sum_n \left( \Delta_D \Psi_{n+1}^\dagger \tau_+ \sigma_0 \Psi_n + \text{h.c.} \right),
\label{Hamiltonian_zero}
\end{gather}
and
\begin{equation}
\mathcal{H}_{\text{defect}} = \Psi_{i_0}^\dagger\delta\hat{h}\Psi_{i_0},
\end{equation}
where $\delta\hat{h}=\Delta_F^0 \tau_0 \sigma_x + \Delta_z^0 \tau_0 \sigma_z$. 

To find the bound state energy induced by this defect, we employ the Green's function formalism. The full Green's function $G^{sd}$ satisfies the Dyson equation, which has the formal solution:
\begin{gather}
G_{ij}^{sd}\left(k_{y}, \varepsilon\right) = G_{ij}^{0}\left(k_{y}, \varepsilon\right) + \nonumber \\
+G_{i i_0}^{0}\left(k_{y}, \varepsilon\right) \delta\hat{h}
\bigl[ 1 - G^{0}\left(k_{y}, \varepsilon\right)\delta\hat{h} \bigr]^{-1} 
G_{i_0 j}^{0}\left(k_{y}, \varepsilon\right).
\label{sd-Green-solution-correct}
\end{gather}
where $G_{ij}^{0}$ is the Green's function of the perfect system (with Hamiltonian~\eqref{Hamiltonian_zero}) and $G^0=G_{ii}^{0}$. The new poles of $G^{sd}$, which correspond to bound state energies, are given by the condition:
\begin{equation}
\det\bigl[1 - G^{0}\left(k_{y}, \varepsilon\right) \delta\hat{h}\bigr]=0.
\label{pole-condition}
\end{equation}

\noindent Inside the bulk gap, the unperturbed Green's function $G^{0}$ is real, allowing Eq.~(\ref{pole-condition}) to have solutions. To find the bound state energy for arbitrary momentum $k_y$, we use the full expression for the local Green's function. For energies inside the bandgap, where $\varepsilon_{\perp}^{2} < (\Delta_{S} - \Delta_{D})^{2}$, it takes the form:
\begin{gather}
G_{ll}^{0}=\sum_{k_{x}}G_{\mathbf{k}}^{0}\to
\int_{-\pi}^{\pi}\frac{dk_{x}}{2\pi}\,
\left(\varepsilon-\mathcal{H}_{\text{bulk}}\left(\mathbf{k}\right)\right)^{-1} =\nonumber\\
=-\frac{(z_{1}\Delta_{D} + \Delta_{S})\tau_{x} + \varepsilon + v_{F} k_{y} \tau_{z} \sigma_{z}}
{2\Delta_{S}\Delta_{D} \sqrt{ \frac{(\varepsilon_{\perp}^{2} - \Delta_{S}^{2} - \Delta_{D}^{2})^{2}}{4\Delta_{S}^{2}\Delta_{D}^{2}} - 1 }},
\label{eq:Green_ll_gap}
\end{gather}
where 
\begin{gather}
z_{1}=\frac{\varepsilon_{\perp}^{2}-\Delta_{S}^{2}-\Delta_{D}^{2}}{2\Delta_{S}\Delta_{D}}
+\left|\sqrt{\frac{(\varepsilon_{\perp}^{2}-\Delta_{S}^{2}-\Delta_{D}^{2})^{2}}{4\Delta_{S}^{2}\Delta_{D}^{2}}-1}\right|,
\end{gather}
and $\varepsilon_{\perp}^{2} = \varepsilon^{2} - v_{F}^{2}k_{y}^{2}$. We can rewrite the Green's function as
\begin{gather}
G_{ll}^{0}=G_0\mathbb{I}_{4\times4}+G_1\tau_x\sigma_0+G_2\tau_z\sigma_z,
\label{eq:Green_ll_gap+}
\end{gather}
Then we have the following equation for in-gap states
\begin{gather}
1+\left(\Delta_{F}^{2}+\Delta_{z}^{2}\right)^{2}\left(G_1^{2}+G_2^{2}-G_0^{2}\right)^{2}-\nonumber\\
-2\left(\Delta_{F}^{2}+\Delta_{z}^{2}\right)\left(G_0^{2}+G_1^{2}\right)+2\left(\Delta_{F}^{2}-\Delta_{z}^{2}\right)G_2^{2}=0.
\label{ingap states}
\end{gather}
The spectral signature of the magnetic defect provides a powerful tool to distinguish between the topological and trivial bulk phases of the host system. Figure~\ref{new states into gap} presents the energy of the resulting in-gap bound states, calculated at $k_y = 0$ from Eq.~\eqref{ingap states}, as a function of the Zeeman splitting strength $\Delta_Z$ for two fixed values of the spin-flip tunneling amplitude $\Delta_F$.

In the topological phase ($\Delta_S = 5$ meV $< \Delta_D = 10$ meV), the defect generates four distinct states within the bulk gap [Fig.~\ref{new states into gap}(a)]. The initially degenerate levels split, and all states exhibit only a slight dependence on the Zeeman splitting. Remarkably, none of these in-gap states cross zero energy for any value of the perturbation, reflecting the protection by the non-trivial topology.

In the trivial phase ($\Delta_S = 10$ meV $> \Delta_D = 5$ meV), the spectrum reveals only two in-gap states [Fig.~\ref{new states into gap}(b)]. The other pair of states remains pinned to the bulk band edges and does not depend on the perturbation. In stark contrast to the topological case, the two in-gap states in the trivial phase do cross zero energy at a critical value of $\Delta_Z$. This distinct behavior of the defect-induced in-gap states provides a sensitive local probe of the global topological order. This difference in the spectral response to a local magnetic perturbation serves as a clear spectroscopic fingerprint of the global topological order (see below).

\begin{figure}
    \centering
    \includegraphics[width=\linewidth]{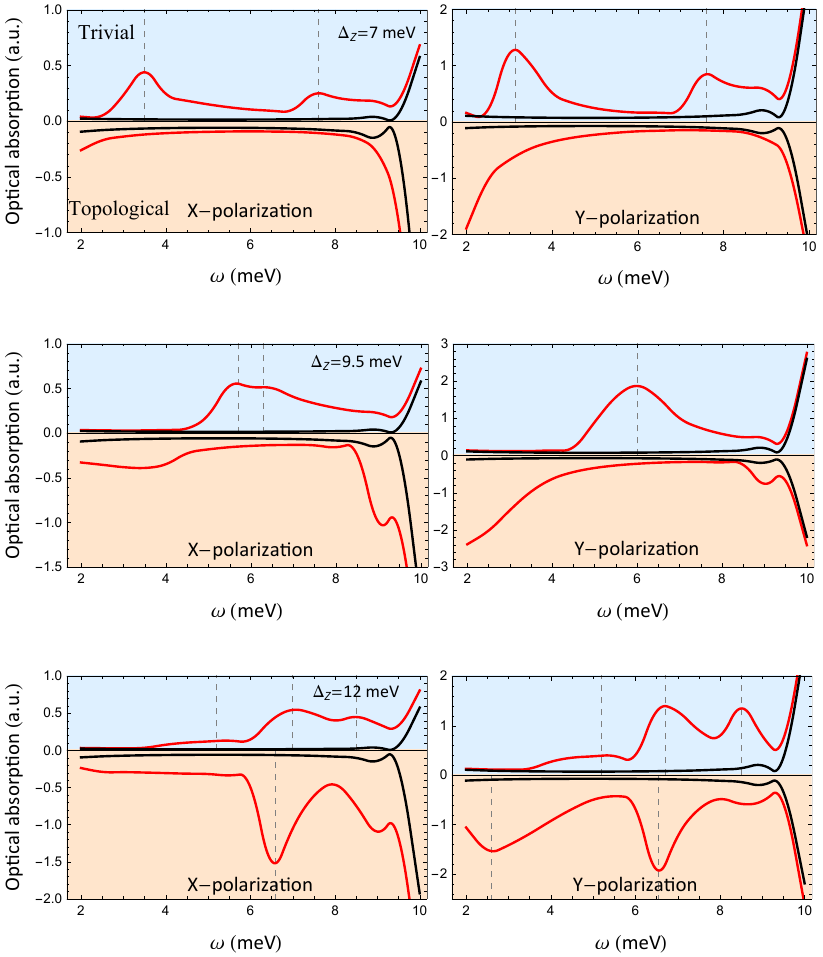}
    \caption{ Optical absorption of a plane electromagnetic wave as a function of frequency in the presence of a defect stripe (red curves) and without the defect (black curves). The left and right columns correspond to $X$- and $Y$-polarization of the incident wave, respectively. The three rows correspond to different values of the Zeeman splitting $\Delta_z = 7$, $9.5$, and $12$ meV (top, middle, and bottom panels). Each panel is divided into two regions: the blue and light orange shaded areas mark the trivial and topological phases, respectively.}
    \label{fig:optical absorption}
\end{figure}

So far, we have employed units in which $\hbar = 1$ and $a_x = a_y = 1$, where $a_x$ and $a_y$ are the lattice constants along the $x$ and $y$ directions, respectively. These parameters are introduced explicitly in the numerical calculations of this section; in the remainder of the paper they were not needed. Furthermore, for the study of $k_y$-dependence and when integrating over the entire Brillouin zone in $k_y$, we perform the standard replacements
\begin{gather}
k_y \to \frac{1}{a_y} \sin(k_y a_y), \nonumber\\
k_y^2 \to \frac{4}{a_y^2}\sin^2\left(\frac{k_y a_y}{2}\right), \nonumber
\end{gather}
which are commonly used in the theory of topological insulators (see, e.g., Ref.~\cite{shen2012topological}).

Figure~\ref{new states density of states} shows the $k_y$-dispersion and the density of states of the in-gap defect state. The density of states exhibits sharp peaks near the defect-state energies in the trivial phase. This behavior is attributed to the quasi-one-dimensional nature of these states: they are localized along the $x$-direction and delocalized along the $y$-direction. By contrast, due to the two-dimensional character of the bulk bands, no pronounced van Hove singularities appear near the conduction band bottom or the valence band top. The arrows in the density-of-states panel indicate the main optical transitions originating from the valence band, which contribute to the optical response.

To further characterise the two phases, we evaluate the optical absorption associated with the in‑gap defect states. We consider a normally incident monochromatic electromagnetic wave with vector potential $\bm{A}(\bm{r},t)=A_0\bm{\eta}\cos(kz-\omega t)$, where $A_0$ and $\bm{\eta}=(\cos\chi,\sin\chi,0)$ are the amplitude and polarisation vector, respectively. Using first‑order time‑dependent perturbation theory, the absorption coefficient is given by~\cite{mahan2013many}
\begin{gather}
\alpha(\eta,\omega)= \frac{\pi e^2}{c\omega\epsilon_0 S}
\sum_{n,n',\bm{k}} \times\nonumber\\ \left| \bigl(\hat{\bm{v}}_{nn'}(\bm{k})\cdot\bm{\eta}\bigr) \right|^2
\bigl[ f_{n',\bm{k}} - f_{n,\bm{k}} \bigr]
\delta\bigl( \varepsilon_{n,\bm{k}} - \varepsilon_{n',\bm{k}} - \hbar\omega \bigr),\label{optical absorption}
\end{gather}
where $\hat{\bm{v}}_{nn'}=\langle n|\hat{\bm{v}}|n'\rangle$ is the mathrix elements of velocity operator, $n=(v,c,d)$ corresponds to valence, conduction, and defect bands, respectively. For the case of low temperature which we consider we have $f_{v,\bm{k}}\approx1,~f_{d(c),\bm{k}}\approx0$. Then 
\begin{gather}
\alpha(\eta,\omega)\approx \frac{\pi e^2}{c\omega\epsilon_0 S}
\sum_{v,c,d,k_x,k_y} \times\nonumber\\ |\langle v| \hat{\bm{v}}\cdot\bm{\eta}|c\rangle|^2
\delta\bigl( \varepsilon_{c,k_x,k_y} - \varepsilon_{v,k_x,k_y} - \hbar\omega \bigr)+\nonumber\\
+|\langle v| \hat{\bm{v}}\cdot\bm{\eta}|d\rangle|^2 
\delta\bigl( \varepsilon_{d,k_y} - \varepsilon_{v,k_x,k_y} - \hbar\omega \bigr).
\label{optical absorption append}
\end{gather}
The summation over the band indices 
$v$ and $c$ already accounts for both spin projections. The explicit forms of the eigenfunctions for the valence, conduction, and defect bands are presented in Appendix~\ref{Appendix C}. The details of the velocity matrix elements and the derivation of the optical absorption formula are provided in Appendix~\ref{Appendix D}.

For the operator of the velocity we have
\begin{gather}
\hat{v}_x=\frac{\Delta_D a_x}{\hbar}\left(\cos\left(k_x a_x\right)\tau_y\sigma_0-\sin\left(k_x a_x\right)\tau_x\sigma_0\right),\\
\hat{v}_y=v_F \cos\left(k_y a_y\right) \tau_z\sigma_z,
\end{gather}

Figure~\ref{fig:optical absorption} presents the calculated optical absorption in the presence of a defect layer (red curves) and without it (black curves). The calculations are performed for both the trivial ($\Delta_S = 10$ meV $> \Delta_D = 5$ meV) and topological ($\Delta_S = 5$ meV $< \Delta_D = 10$ meV) phases, and for two polarizations of the incident wave. A clear distinction in the optical response between the two phases is observed, arising from the different behavior of the in-gap defect states and the corresponding matrix elements. First, in contrast to the topological phase, the trivial phase exhibits regions of zero absorption. Second, in the trivial phase two absorption peaks appear at frequencies $\omega_{\nu1}$ and $\omega_{\nu2}$ (see Fig.~\ref{new states into gap}). These peaks approach each other as $\Delta_Z$ approaches $\Delta_S$ (indicated by the red arrow in Fig.~\ref{new states into gap}), and move apart again upon further increase of $\Delta_Z$. In the topological phase, absorption peaks emerge only for relatively large values of $\Delta_Z$, which is attributed to the specific behavior of the transition matrix elements. Thus the optical absorption provides an independent, experimentally accessible fingerprint of the topological order, complementing the static spectral features discussed above.

\begin{figure}
    \centering
    \includegraphics[width=\linewidth]{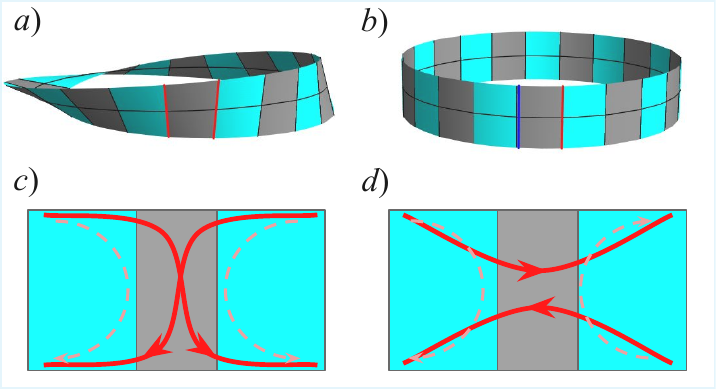}
    \caption{(Color online) A comparison of two distinct global topologies arising from closing a quasi-1D ribbon. (a) M\"obius strip geometry, formed by connecting the ends of the ribbon with a half-twist. (b) Cylindrical geometry, formed by direct connection of the ends. The fundamental difference manifests in the boundary conditions for the chiral edge modes at the junction (highlighted in red). (c) In the M\"obius strip, the connection forces edge modes of the \textit{same} chirality to meet at the seam, leading to their hybridization and intersection in real space. (d) In the cylinder, the connection brings together edge modes of \textit{opposite} chirality, as is the case at all other interfaces in the array. This results in two spatially separated edge modes, localized on opposite physical boundaries and propagating without intersection.}
    \label{Mobius}
\end{figure}

\section{Discussion and perspectives}

In summary, we have developed a comprehensive theory for a quasi-1D magnetic heterostructure, mapping it onto a novel hybrid SSH-Shockley model. Our symmetry analysis unequivocally classifies the system in class AIII, characterized by a chiral symmetry and a $\mathbb{Z}$ topological invariant. The calculated phase diagram reveals a rich structure, including a remarkable phase where a null global invariant masks the existence of robust, spatially separated end states—a direct manifestation of compensated non-trivial topology from opposing subsystem winding numbers. Furthermore, we demonstrated that the in-gap spectrum induced by a single magnetic defect provides a distinct spectroscopic fingerprint of the global topological order. It would be interesting to explore how external electric and magnetic fields modify the defect states and their optical response, following earlier studies in Dirac systems~\cite{Alisultanov2016,Alisultanov2017}.

This work establishes a versatile and tunable platform for exploring higher-order topological phenomena through material engineering and magnetic patterning. The uncovered principles of hidden topology and local probing are general and could be extended to other topological material classes.

\begin{figure}
    \centering
    \includegraphics[width=\linewidth]{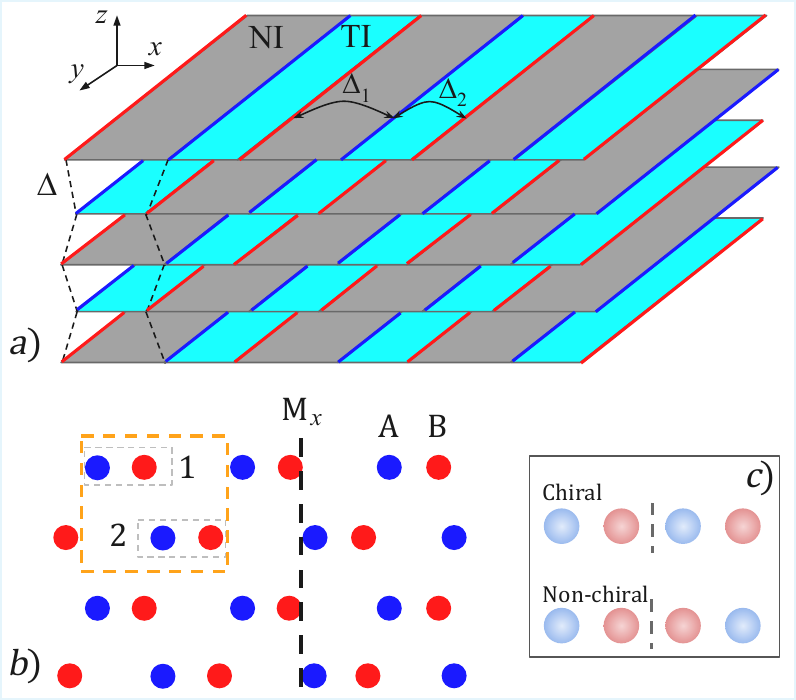}
    \caption{(Color online) Layered heterostructure and its emergent topology. (a) A three-dimensional stack of the proposed quasi-1D planar ribbons along the $Z$-axis. (b) An equivalent planar representation of the layered structure, illustrating the two-dimensional periodicity in the $X$-$Z$ plane. This system possesses a non-symmorphic projective translational symmetry $\mathrm{L}_z$, while translational invariance along $X$ is explicitly broken, rendering the system an insulator. (c) Due to the nontrivial action of the $\mathrm{L}_z$ symmetry, the projection of the Brillouin zone onto the $XZ$-plane yields a spectrum with edge states characterized by a M\"obius strip topology. Further compactification of the Brillouin zone along the $Y$-direction results in a manifold with the topology of a Klein bottle.}
    \label{multilayer Mobius}
\end{figure}

Furthermore, the proposed platform, being free of magnetic impurities, provides an interesting opportunity to investigate nontrivial symmetry-protected and topological phenomena. Specifically, the nanoribbon can be closed into two distinct geometries: a cylinder or a Möbius strip (see Fig.~\ref{Mobius}). As the figure illustrates, these two topologies have profoundly different consequences for the edge modes. In the cylindrical geometry, the counter-propagating edge modes remain spatially separated and localized on opposite boundaries. In contrast, in the Möbius strip geometry, the edge modes are forced to hybridize at the "seam" where the ends are connected, forming an analogue of a Hopf bundle. This fundamental difference arises because the connection in the cylinder brings together edges with opposite chiralities, whereas the Möbius strip connection identifies edges with the same chirality, leading to a direct intersection of the modes (see Fig.~\ref{Mobius}). This twisting effectively alters the boundary conditions. The Hamiltonian for the Möbius strip thus differs from that of the cylinder by the presence of a single topological defect (TD) at the junction, characterized by the identification of similar chiralities instead of opposing ones. In terms of projective translation symmetry, this defect breaks the $L_x \otimes G$ symmetry mentioned above (see text before~\eqref{eq:spectrum}). We can express this Hamiltonian as:
\begin{equation}
\mathcal{H} = \mathcal{H}_{\text{bulk}} + \mathcal{H}_{\text{TD}}
\end{equation}
with
\begin{equation}
\mathcal{H}_{\text{TD}} = \Psi_{N}^\dagger v_F k_y \tau_z\otimes \left(\sigma_0-\sigma_z\right)\Psi_{N},
\end{equation}
where $N$ is number of the junction site and $\mathcal{H}_{\text{bulk}}$ is given by~\eqref{Hamiltonian_zero}. This defect is topological and cannot be eliminated continuously.

Another nontrivial topological feature emerges when considering the heterostructure depicted in Fig.~\ref{multilayer Mobius}. In this case, an additional degree of freedom—the layer index along the Z-axis—comes into play. The Hamiltonian for such a structure can be formulated as follows:
\begin{gather}
\mathcal{H}(\mathbf{k})\!=\! k_y \gamma_0\otimes\tau_0\otimes\tau_z\otimes \sigma_z\!+\nonumber\\
+\!\gamma_0\otimes\tau_0\otimes\begin{pmatrix}\!
    0&\Delta_1\!+\!\Delta_2e^{-ik_x}\\
    \Delta_1\!+\!\Delta_2e^{ik_x}&0\!
\end{pmatrix}
\otimes\sigma_0
\!+\nonumber\\
+\Delta\left(1+e^{-ik_z}\right)\gamma_+\otimes\tau_+\otimes\begin{pmatrix}
    0&e^{-ik_x}\\
    1&0
\end{pmatrix}\otimes\sigma_0+\nonumber\\
+\Delta\left(1+e^{ik_z}\right)\gamma_-\otimes\tau_-\otimes\begin{pmatrix}
    0&1\\
    e^{ik_x}&0
\end{pmatrix}\otimes\sigma_0,
\label{nonsym Hamiltonian momentum}
\end{gather}
where $\bm{\gamma}=(\gamma_x,\gamma_y,\gamma_z)$ are Pauli matrices in the basis (1,2) of displacements along the $Z$ axis (see Fig.~\ref{multilayer Mobius}), $\gamma_{\pm}=1/2(\gamma_x\pm i\gamma_y)$, and the matrix products are tensor products (the direct product symbol is omitted for brevity). We put $\upsilon_F=1$. It is straightforward to show that such a Hamiltonian remains invariant under the transformation
 $\mathrm{L}_z=\begin{pmatrix}
    0&1\\
    e^{ik_z}&0
\end{pmatrix}\otimes\tau_x\otimes\tau_x\otimes\sigma_x$:
\begin{gather}
\mathrm{L}_z^{-1}\mathcal{H}(-k_x,k_y,k_z)\mathrm{L}_z=\mathcal{H}(k_x,k_y,k_z).
\end{gather}
The operator $\mathrm{L}_z$ corresponds to a translation along the $Z$-axis combined with a reflection in the $YZ$-plane. Indeed, this operation reverses the spin projection, as well as transforming $k_x \to -k_x$. Such Hamiltonian has chiral symmetry with the operator
$\hat{C}=\gamma_0\otimes\tau_z\otimes\sigma_x$: $\{\hat{C},\mathcal{H}\}=0$.

The transformation $\mathrm{L}_z$ is related to projective non-symmorphic crystal symmetries, which have become an important direction in modern research. Such symmetries arise in the presence of gauge fields, for instance, in the context of $\mathbb{Z}_2$ lattice gauge theories, spin liquids, or artificial crystals~\cite{zhao2020z,chen2022brillouin,zhang2023general,zhang2025brillouin}. $\mathbb{Z}_2$ projective translational symmetry gives rise to fundamentally new topological phenomena~\cite{zhao2020z}. In particular, the anticommutation relation between translation operators in such a setting induces twofold band degeneracy and precludes a global trivialization of the spinor structure in momentum space~\cite{zhao2020z}. Combining this symmetry with time-reversal symmetry yields fourfold-degenerate Dirac points, while its partial breaking leads to a Möbius insulator phase characterized by distinctive edge states~\cite{zhao2020z}.

Since $\mathrm{L}_z$ is a symmetry, its eigenfunctions can be used to diagonalize the Hamiltonian matrix at planes $k_x=0,\pi$. It is easy to show that the following matrix diagonalizes the operator $\mathrm{L}_z$ with eigenvalues $\pm e^{ik_z/2}$: 
\begin{gather}
U\left(k_z\right)=T_z\otimes U_{\tau}\otimes U_{\sigma},
\end{gather}
where 
\begin{gather}
T_z=\frac{1}{\sqrt{2}}\begin{pmatrix}
    1&1\\
    e^{ik_z/2}&-e^{ik_z/2}
\end{pmatrix},\\
U_{\tau}=U_{\sigma}=\frac{1}{\sqrt{2}}\begin{pmatrix}
    1&1\\
    1&-1
\end{pmatrix},
\end{gather}
After diagonalization we have following 
\begin{gather}
U^{\dagger}\mathcal{H}(k_0,k_y,k_z)U=\begin{pmatrix}
    h_1\left(k_y,k_z\right)&0\\
    0&h_2\left(k_y,k_z\right)
\end{pmatrix},
\end{gather}
where $k_0=0,\pi$ and 
\begin{gather}
h_{1,2}\left(k_y,k_z\right)=k_y\tau_x\sigma_x
+\left(m\pm2\Delta \cos\frac{k_z}{2}\right)\tau_z\sigma_0,
\end{gather}
where $m=\Delta_1\pm\Delta_2$ at $k_x=0,\pi$. We note that 
\begin{gather}
\tau_x\sigma_z h_{1,2}\tau_x\sigma_z=-h_{1,2},\\
h_{1,2}\left(k_y,k_z+2\pi\right)=h_{2,1}\left(k_y,k_z\right).
\end{gather}

As is evident from these expressions, the Hamiltonian exhibits nonperiodic behavior in momentum space along the $k_z$ direction, which directly manifests itself in the emergence of Möbius topology in the electronic band structure. Indeed, the explicit form yields the relation $\mathcal{H}\left(k_z+2\pi\right)=\gamma_x\mathcal{H}\left(k_z\right)\gamma_x$, which is a similarity transformation that leaves the spectrum invariant but rotates the basis of eigenfunctions. Consequently, the eigenstates satisfy $|\Psi\rangle\left(k_z+2\pi\right)=\gamma_x |\Psi\rangle\left(k_z\right)$, from which, in the block-diagonal representation, we obtain $|\psi_{1,2}\rangle\left(k_z+2\pi\right)=|\psi_{2,1}\rangle\left(k_z\right)$, where $|\psi_{1,2}\rangle$ are the eigenfunctions of the operators $h_{1,2}$. This structure is a direct consequence of the nonsymmorphic symmetry of the system and serves as a key signature of the realization of Möbius topology in the band spectrum.

\begin{figure}[t]
    \centering
    \includegraphics[width=0.8\linewidth]{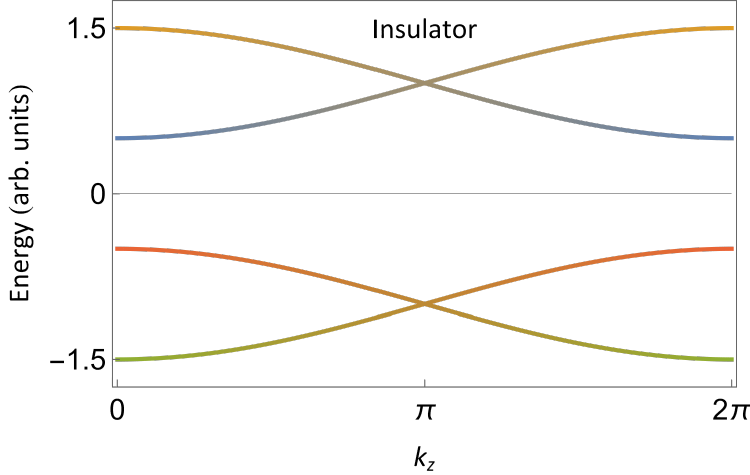}
    \includegraphics[width=0.8\linewidth]{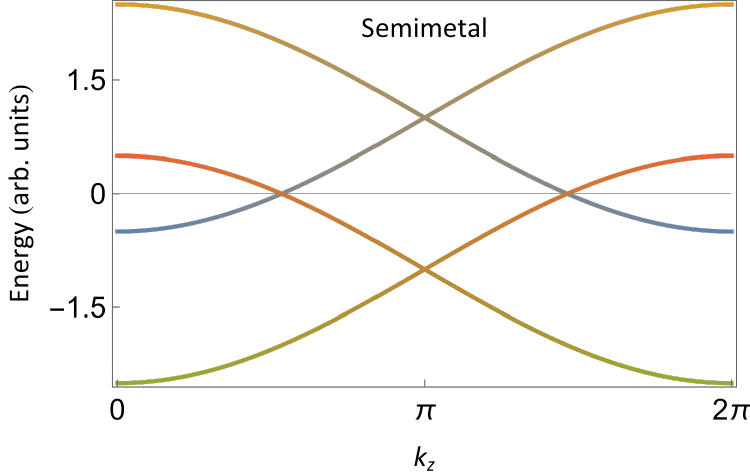}
    \caption{Band structure exhibiting Möbius topology for the gapped ($m>2\Delta$, top panel) and gapless ($m<2\Delta$, bottom panel) phases. The two band branches interchange upon traversing the Brillouin zone; returning to the initial point requires a double loop. The spectra are shown at $k_y=0$.}
    \label{fig:Mobius}
\end{figure}

Focusing on the point $k_y = 0$, the energy spectrum takes the analytically tractable form
\begin{gather}
e_{1,2}=\pm\left(m\pm2\Delta \cos\frac{k_z}{2}\right),
\end{gather}
where the interplay of parameters dictates the topological phase. When $m > 2\Delta$, the spectrum is gapped. Notably, the eigenstates within each band (conduction and valence) form distinct M\"obius pairs, as illustrated schematically in Fig.~\ref{fig:Mobius}(a). A topological phase transition occurs at the critical point $m = 2\Delta$. For $m < 2\Delta$, the system becomes gapless, hosting a semimetallic phase. In this regime, the previously separated M\"obius pairs intersect, resulting in the formation of symmetry-protected Dirac points at zero energy, shown in Fig.~\ref{fig:Mobius} (b).

A comprehensive phase diagram emerges from this analysis, governed by the parameters $\Delta_1$, $\Delta_2$, and $\Delta$ at the high-symmetry points $k_x = 0, \pi$: 1) Fully gapped Phase: When $\Delta_1 \pm \Delta_2 > 2\Delta$ holds for both signs, the system is gapped at both $k_x=0$ and $k_x=\pi$. 2) Fully gapless Phase: When $\Delta_1 \pm \Delta_2 < 2\Delta$ holds for both signs, the system is gapless at both $k_x=0$ and $k_x=\pi$, hosting a total of four Dirac nodes at the Fermi level. 3) Mixed Phase: When $\Delta_1 - \Delta_2 < 2\Delta < \Delta_1 + \Delta_2$, a hybrid phase is realized. The spectrum gapped at $k_x=0$ but gapless at $k_x=\pi$, yielding two Dirac points at zero energy. Transitions between these phases can be achieved by changing $\Delta$, for example, by uniaxial deformation along Z.

\textit{Acknowledgments}. The study is supported by the Ministry of Science and Higher Education of the Russian Federation (Goszadaniye) (No. FSMG-2026-0012). The study of the problem of a single magnetic defect and the spectroscopic features has been supported by the Russian Science Foundation (Project No. 26-12-00345).

\appendix

\section{Physical realization of the alternating TI/NI stripe structure in twisted bilayer TMDs}
\label{Appendix A}

In this Appendix we argue that the alternating topological insulator (TI)/normal insulator (NI) stripe geometry introduced in Fig.~\ref{fig:skt} of the main text can be realistically achieved in twisted bilayers of transition metal dichalcogenides (TMDs) from the 1T' structural family, including $\text{MoTe}_2$, $\text{WSe}_2$, etc. The argument combines recent experimental observations, first‑principles calculations, and the Jackiw‑Rebbi mechanism.

It has been demonstrated experimentally that twisted bilayers of $\text{WTe}_2$ at twist angles of approximately $58^\circ$ and $62^\circ$ form purely one‑dimensional moiré patterns consisting of parallel stripes with long‑range periodicity only along one direction~\cite{yang2025intrinsic}. Similar behaviour can be expected in other TMDs with analogous crystal structures, such as $\text{MoTe}_2$ and $\text{WSe}_2$, because their moiré physics is governed by the same geometric principles. Thus the quasi‑one‑dimensional geometry required for our model is already present in real samples.

A single monolayer of $1T'-\text{TMD}$ is a well‑known two‑dimensional topological insulator. The band structure exhibits a band inversion with an energy scale $\Delta_{\text{inv}}$ of the order of $200-300$~meV, as established by first‑principles calculations~\cite{qian2014quantum}. This inverted gap is responsible for the topological (quantum spin Hall) state. Importantly, it has also been shown that a perpendicular electric field (gate voltage) can modify $\Delta_{\text{inv}}$ and eventually close it, driving a transition to a trivial insulator~\cite{qian2014quantum}. Consequently, the "TI segments" in our Fig.~\ref{fig:skt} correspond to regions where the bilayer remains topologically nontrivial, while "NI segments" correspond to regions where the effective mass gap has been suppressed or the band ordering has returned to normal.

Within the moiré pattern, different stripes correspond to different local stacking registries of the two monolayers. This has been analysed in detail using density functional theory for twisted $\text{WTe}_2$~\cite{magorrian2024strain}. Depending on the local stacking, the interlayer hopping integral \(t_{\perp}\) varies strongly — it can be large, small, or even vanish, and it may change sign. Under moderate strain (of the order of one percent) the confining potential between stripes reaches an amplitude of more than 200 meV, which is comparable to the intrinsic band inversion scale \(\Delta_{\text{inv}}\)~\cite{magorrian2024strain}. Therefore different stripes in the same twisted bilayer can have drastically different interlayer coupling strengths.

The crucial physical mechanism is the competition between interlayer coupling and the intrinsic band inversion. When the interlayer hopping \(t_{\perp}\) exceeds \(\Delta_{\text{inv}}\), the hybridization between the two monolayers overcomes the inversion and the bilayer becomes a trivial insulator. When \(t_{\perp} < \Delta_{\text{inv}}\), the bilayer remains topological. DFT calculations have explicitly shown that the Fermi surface topology changes qualitatively depending on the local stacking: some stackings (weak coupling) preserve the topological character, while others (strong coupling) become trivial~\cite{magorrian2024strain}. As a result, different stripes within the same twisted bilayer are topologically distinct — some are TI, some are NI. Because the one‑dimensional moiré pattern consists of alternating stripes with different stackings, the alternating TI/NI structure of Fig.~\ref{fig:skt} emerges spontaneously, without any external lithographic patterning.

At each interface between a TI stripe and an NI stripe the effective mass term changes sign, and according to the Jackiw‑Rebbi mechanism a one‑dimensional chiral mode is bound to the interface and propagates along the stripe direction (the \(y\)-axis in our model)~\cite{semenoff2008domain}. Neighbouring chiral modes — those on opposite sides of a TI stripe and those separated by an NI stripe — hybridise via quantum tunnelling (see Appendix~\ref{Appendix B}). This tunnelling gives rise to two independent amplitudes: \(\Delta_S\) (tunnelling across a TI stripe, intra‑cell hopping) and \(\Delta_D\) (tunnelling across an NI stripe, inter‑cell hopping). These amplitudes are precisely the parameters that enter the tight‑binding Hamiltonian~\eqref{Hamiltonian} of the main text. 

Thus, the proposed alternating TI/NI stripe structure is directly realisable in twisted bilayer TMDs. Our model therefore provides a novel theoretical framework for describing the low‑energy physics and topological properties of these experimentally accessible systems, while going beyond previous analyses by explicitly computing the winding number \(\nu = 0,1,2\), predicting a spectroscopic fingerprint via a magnetic defect.

\section{Tunneling between edge modes in two-dimensional topological insulator}
\label{Appendix B}

In this Appendix, we briefly review the mechanism of tunneling between helical edge states in a finite-width strip of a two-dimensional topological insulator. The results are based on the continuum Bernevig-Hughes-Zhang (BHZ) model and its analytical solution for a strip geometry (this problem has been studied in~\cite{zhou2008finite} and here we show the main results of this pioneering work). Our goal is to demonstrate that the edge states localized on opposite edges of the strip hybridize, and that this hybridization occurs between modes with the same spin index  (they belong to the same block of the BHZ Hamiltonian) but opposite chiralities (helicities). The overlap of their wave functions at finite width \(L\) leads to a finite energy gap and provides the microscopic origin of the effective tunneling parameters of our model used in the main text.

The BHZ Hamiltonian for quantum wells is given by (see~\cite{bernevig2006quantum}):

\begin{gather}
\mathcal{H}(\mathbf{k}) = \begin{pmatrix} H(\mathbf{k}) & 0 \\ 0 & H^{*}(-\mathbf{k}) \end{pmatrix}, \\ H(\mathbf{k}) = \epsilon_{\mathbf{k}} + d^{a}(\mathbf{k})\sigma^{a},
\end{gather}
where \(\sigma^{a}\) are Pauli matrices acting on the orbital basis \((E1, H1)\) (electron-like and hole-like states). The explicit terms are:
\begin{gather}
\epsilon_{\mathbf{k}} = C - D(k_x^2 + k_y^2), \\ 
d^1 = A k_x, \quad d^2 = A k_y, \\
d^3 = M - B(k_x^2 + k_y^2).
\end{gather}
The parameters \(A, B, C, D, M\) are material-specific. In the inverted regime (quantum spin Hall phase), \(M/B > 0\). The upper block \(H(\mathbf{k})\) is conventionally associated with spin-up electrons, and the lower block with spin-down electrons. 

We consider a strip of width \(L\) with periodic boundary conditions along \(x\) and open boundary conditions along \(y\). The momentum \(k_x\) is conserved, and we replace \(k_y \to -i\partial_y\). The eigenproblem for the upper block is:
\begin{gather}
\big[ M - B_+(k_x^2 - \partial_y^2) \big]\psi_1 + A(k_x - \partial_y)\psi_2 = E\psi_1,
\end{gather}
\begin{gather}
A(k_x + \partial_y)\psi_1 - \big[ M - B_-(k_x^2 - \partial_y^2) \big]\psi_2 = E\psi_2,
\end{gather}
with \(B_{\pm} = B \pm D\), and \(\Psi_{\uparrow} = e^{ik_x x}(\psi_1(y), \psi_2(y))^T\). The solutions for the lower block are obtained by time-reversal: \(\Psi_{\downarrow} = \Theta \Psi_{\uparrow}\) with \(\Theta = -i\sigma_y K\).

Assuming \(\psi_{1,2} \propto e^{\lambda y}\), the characteristic equation yields four roots \(\pm\lambda_1, \pm\lambda_2\) with:
\begin{gather}
\lambda_{1,2}^2 = k_x^2 + F \pm \sqrt{F^2 - \frac{M^2 - E^2}{B_+ B_-}}, \\ 
F = \frac{A^2 - 2(MB + ED)}{2B_+ B_-}.
\end{gather}

For parameters corresponding to the inverted regime, \(\lambda_{1,2}\) are real for small \(k_x\) and describe edge states localized near \(y = \pm L/2\).  The secular equation for the energies \(E(k_x)\) at finite \(L\) gives a finite gap at \(k_x = 0\):
\[
\Delta \equiv E_+ - E_- \simeq \frac{4|A B_+ B_- M|}{\sqrt{B^3 (A^2 B - 4B_+ B_- M)}} \, e^{-\lambda_2 L},
\]
which decays exponentially with \(L\) because \(\lambda_2^{-1}\) is the characteristic decay length of the edge states into the bulk.

In the limit of large \(L\) (but still finite), the wave functions simplify. One finds that the ratio \(\psi_2 / \psi_1\) approaches constants near the two edges:

\[
\frac{\psi_2}{\psi_1} \to \pm \eta,\qquad \eta = \frac{B_+(\lambda_1 + \lambda_2)}{A}.
\]

Thus, the solutions for the upper block near the left and right edges are:

\[
\Psi_{\uparrow, \text{left}} \propto \begin{pmatrix} 1 \\ -\eta \end{pmatrix},\qquad
\Psi_{\uparrow, \text{right}} \propto \begin{pmatrix} 1 \\ +\eta \end{pmatrix}.
\]

Both solutions belong to the same spin block (the upper block). However, they correspond to opposite helicities (chiralities) because the sign of \(\eta\) determines the projection of the physical spin onto the direction of motion. At finite \(L\), the two edge states overlap in space, and the coupling between them is described by the matrix element \(t \sim \langle \Psi_{\uparrow,\text{left}} | \mathcal{H} | \Psi_{\uparrow,\text{right}} \rangle \propto e^{-\lambda_2 L}\). This coupling opens the energy gap \(\Delta = 2|t|\). The tunneling is allowed despite the fact that the two states belong to the same spin block because they have opposite helicity. This mechanism provides the microscopic origin of the tunneling parameters \(\Delta_S\sigma_0\) and \(\Delta_D\sigma_0\) used in our tight-binding model: they are proportional to the overlap integrals between edge states on adjacent TI and NI segments. The key physical insight is that tunneling occurs between states of the same spin but opposite chirality, a unique feature of helical edge states in quantum spin Hall systems.

\section{Eigenfunctions of in-gap bound states induced by a single magnetic stripe}
\label{Appendix C}

In this Appendix we derive the wave functions of the in‑gap states bound to a single magnetic defect. The defect is assumed to be localized at the unit cell \(n=l\) and is described by the operator \(\hat{h}\) acting only on that cell. The unperturbed system is translationally invariant along the \(x\)-direction (the stripe periodicity direction) and along the \(y\)-direction (the edge direction).

The key relation follows from the Dyson equation. The wave function of a bound state with energy \(\varepsilon\) (measured from the Dirac point) and transverse momentum \(k_y\) satisfies

\begin{gather}
\langle n|\psi_d\rangle = G^0_{nl}(\varepsilon,k_y) \hat{h} \langle l|\psi_d\rangle,\label{defect vector}
\end{gather}
where \(G^0_{nl}(\varepsilon,k_y)\) is the retarded Green’s function of the perfect system, and \(\langle l|\psi_d\rangle\) is the wave function of the defect unit cell ($l$-th stripe). For the Green function we set 
$G_{nl}^{0}=G_{ll}^{0}\,e^{-a(\varepsilon,\mathbf{k}_{\perp})|n-l|}$, where
\begin{gather}
a=\frac{1}{\ln\Bigl|\dfrac{\varepsilon_{\perp}^{2}-\Delta_{S}^{2}-\Delta_{D}^{2}}{2\Delta_{S}\Delta_{D}}
      -\Bigl|\sqrt{\dfrac{(\varepsilon_{\perp}^{2}-\Delta_{S}^{2}-\Delta_{D}^{2})^{2}}
                         {4\Delta_{S}^{2}\Delta_{D}^{2}}-1}\Bigr|\Bigr|}.
\end{gather}
The Eq.~\eqref{defect vector} can be obtained from the Schrödinger equation
\begin{gather}
\left(H_0+|l\rangle\hat{h}\langle l|\right) |\psi_d\rangle=\varepsilon |\psi_d\rangle,
\end{gather}
which gives
\begin{gather}
|\psi_d\rangle=\hat{G}^0\left(\varepsilon,k_y\right)|l\rangle \hat{h}\langle l|\psi_d\rangle
\end{gather}
leading to Eq.~\eqref{defect vector}. 

From Eq.~\eqref{defect vector} we can find the defect wave function $\langle l|\psi_d\rangle$ from
\begin{gather}
G^0_{ll}(\varepsilon,k_y) \hat{h} \langle l|\psi_d\rangle = \langle l|\psi_d\rangle.\label{defect vector problem}
\end{gather}

Thus, we have to solve the eigenfunction problem~\eqref{defect vector problem} with the following operators (here we consider the case $\Delta_F=0$)
\begin{gather}
G^0_{ll}=G_0\mathbb{I}_{4\times4}+G_1\tau_x\sigma_0+G_2\tau_z\sigma_z,\\
\hat{h}=\Delta_z \tau_0 \sigma_z.
\end{gather}
Then we can rewrite Eq.~\eqref{defect vector problem} in the following form
\begin{gather}
\left(G_0\tau_0 \sigma_z+G_1 \tau_x\sigma_z+G_2 \tau_z\sigma_0\right) \langle l|\psi_d\rangle = \frac{1}{\Delta_z} \langle l|\psi_d\rangle.
\end{gather}
This equation is understood as an eigenvalue problem where $\langle l|\psi_d\rangle$ is an eigenvector of the matrix on the left‑hand side. Then we have 
\begin{gather}
|\Psi_d\rangle_i=\langle l|\psi_d\rangle_i e^{ik_y y},\\
\langle l|\psi_d\rangle_1=N_{-}\left(0,\frac{\sqrt{G_1^2+G_2^2}-G_2}{G_1},0,1\right)^T,\\
\langle l|\psi_d\rangle_2=N_{-}\left(-\frac{\sqrt{G_1^2+G_2^2}-G_2}{G_1},0,1,0\right)^T,\\
\langle l|\psi_d\rangle_3=N_+\left(0,-\frac{\sqrt{G_1^2+G_2^2}+G_2}{G_1},0,1\right)^T,\\
\langle l|\psi_d\rangle_4=N_+\left(\frac{\sqrt{G_1^2+G_2^2}+G_2}{G_1},0,1,0\right)^T,
\end{gather}
where $N_{\pm}=\frac{1}{\sqrt{1+\left|\frac{\sqrt{G_1^2+G_2^2}\pm G_2}{G_1}\right|^2}}$. 
For eigenvalues we have the following equation
\begin{gather}
1+\Delta_{z}^{4}\left(G_1^{2}+G_2^{2}-G_0^{2}\right)^{2}
-2\Delta_{z}^{2}\left(G_0^{2}+G_1^{2}+G_2^{2}\right)=0.
\label{ingap states with zeeman}
\end{gather}
This equation can be rewritten in the following form
\begin{gather}
\left(G_0-\frac{1}{\Delta_z}\pm R\right)\left(G_0+\frac{1}{\Delta_z}\pm R\right)=0,\label{defect energy}
\end{gather}
where $R^2=G_1^{2}+G_2^{2}$. 

To identify which eigenfunction corresponds to a given eigenvalue, we decompose the defect wave function $\langle l|\psi_d\rangle$ as a direct product of the edge and spin sectors: $|\phi_\tau\rangle\otimes|\phi_\sigma\rangle = (u_\tau, v_\tau)^T \otimes (u_\sigma, v_\sigma)^T$. Then it is clear that $\langle l|\psi_d\rangle_1$ and $\langle l|\psi_d\rangle_3$ correspond to spin down ($|\phi_\sigma\rangle = (0,1)^T$) and to the eigenvalues in the second bracket of~\eqref{defect energy}, while $\langle l|\psi_d\rangle_2$ and $\langle l|\psi_d\rangle_4$ correspond to spin up ($|\phi_\sigma\rangle = (1,0)^T$) and to the eigenvalues in the first bracket of~\eqref{defect energy}. It can be shown (numerically) that in trivial phase only two states appear inside the band gap. These states correspond to the equations $G_0-1/\Delta_z-R=0,~G_0+1/\Delta_z+R=0$ at $\Delta_z>0$ and $G_0-1/\Delta_z+R=0,~G_0+1/\Delta_z-R=0$ at $\Delta_z<0$. Correspondingly, at $\Delta_z>0$ we have in-gap states $\langle l|\psi_d\rangle_1,~\langle l|\psi_d\rangle_4$, but at $\Delta_z<0$ we have in-gap states $\langle l|\psi_d\rangle_3,~\langle l|\psi_d\rangle_2$. In topological phase, we have two in-gap states ($\langle l|\psi_d\rangle_1,~\langle l|\psi_d\rangle_4$ at $\Delta_z>0$) at $|\Delta_z|<\Delta_D$ and four at $|\Delta_z|>\Delta_D$. 

For the calculation of matrix elements, we also need the eigenfunctions of the unperturbed Hamiltonian of our system 
\begin{gather}
\mathcal{H}_0(\mathbf{k}) = v_F k_y \tau_z \sigma_z +\nonumber\\
+ \left( \Delta_S + \Delta_D \cos k_x \right) \tau_x \sigma_0+\Delta_D \sin k_x \tau_y \sigma_0.
\label{zero Hamiltonian}
\end{gather}
The normalized eigenvectors of this Hamiltonian take the following forms
\begin{gather}
|\Psi_0\rangle_i=|v,c\rangle_{1,2} e^{i(k_x x+k_y y)},\\
|v1\rangle=\sqrt{\frac{\varepsilon_k-v_F k_y}{2 \varepsilon _k}} \left(0,-\frac{\varepsilon _k+v_F k_y}{\Delta_S+\Delta_D e^{i k_x}},0,1\right)^T,\\
|v2\rangle=\sqrt{\frac{\varepsilon_k+v_F k_y}{2 \varepsilon _k}} \left(-\frac{\varepsilon _k-v_F k_y}{\Delta_S+\Delta_D e^{i k_x}},0,1,0\right)^T,\\
|c1\rangle=\sqrt{\frac{\varepsilon_k+v_F k_y}{2 \varepsilon _k}} \left(0,\frac{\varepsilon_k-v_F k_y}{\Delta_S+\Delta_D e^{i k_x}},0,1\right)^T,\\
|c2\rangle=\sqrt{\frac{\varepsilon_k-v_F k_y}{2 \varepsilon_k}} \left(\frac{\varepsilon_k+v_F k_y}{\Delta_S+\Delta_D e^{i k_x}},0,1,0\right)^T,\\
\varepsilon_k=\sqrt{\Delta _D^2+\Delta _S^2+2 \Delta _D \Delta _S \cos k_x+v_F^2 k_y^2}.
\end{gather}

\section{Expression for the optical absorption}
\label{Appendix D}

For calculating of optical absorption we use the Eq.~\eqref{optical absorption} in the main text. In general, we have $|n\rangle=\left(|\Psi_0\rangle_i,~|\Psi_d\rangle_i\right)$. Taking into account that the $\omega>0$ we can conclude that $\varepsilon_{n,\bm{k}}>\varepsilon_{n',\bm{k}}$. Then $f_{n',\bm{k}}- f_{n,\bm{k}}>0$ and
\begin{gather}
\alpha(\eta,\omega)= \frac{\pi e^2}{c\omega\epsilon_0 S}
\sum_{v,c=1,2,d=1-4,\bm{k}} \times\nonumber\\ |\langle v| \hat{\bm{v}}\cdot\bm{\eta}|c\rangle|^2 \left(f_{v}- f_{c}\right)
\delta\bigl( \varepsilon_{c,\bm{k}} - \varepsilon_{v,\bm{k}} - \hbar\omega \bigr)+\nonumber\\
+|\langle v| \hat{\bm{v}}\cdot\bm{\eta}|d\rangle|^2 \left(f_{v}- f_{d}\right)
\delta\bigl( \varepsilon_{d,\bm{k}} - \varepsilon_{v,\bm{k}} - \hbar\omega \bigr)+\nonumber\\
+|\langle d| \hat{\bm{v}}\cdot\bm{\eta}|d'\rangle|^2 \left(f_{d'}- f_{d}\right)
\delta\bigl( \varepsilon_{d,\bm{k}} - \varepsilon_{d',\bm{k}} - \hbar\omega \bigr)+\nonumber\\
+|\langle d| \hat{\bm{v}}\cdot\bm{\eta}|c\rangle|^2 \left(f_{d}- f_{c}\right)
\delta\bigl( \varepsilon_{c,\bm{k}} - \varepsilon_{d,\bm{k}} - \hbar\omega \bigr),
\label{optical absorption appendix}
\end{gather}
For the case of low temperature which we consider we have $f_{v,\bm{k}}\approx1,~f_{d(c),\bm{k}}\approx0$. Then 
\begin{gather}
\alpha(\eta,\omega)\approx \frac{\pi e^2}{c\omega\epsilon_0 S}
\sum_{v,c,d,k_x,k_y} \times\nonumber\\ |\langle v| \hat{\bm{v}}\cdot\bm{\eta}|c\rangle|^2
\delta\bigl( \varepsilon_{c,k_x,k_y} - \varepsilon_{v,k_x,k_y} - \hbar\omega \bigr)+\nonumber\\
+|\langle v| \hat{\bm{v}}\cdot\bm{\eta}|d\rangle|^2 
\delta\bigl( \varepsilon_{d,k_y} - \varepsilon_{v,k_x,k_y} - \hbar\omega \bigr).
\label{optical absorption append}
\end{gather}

The summation over the band indices 
$v$ and $c$ already accounts for both spin projections. We used this formula in the main text. 

Finally, let us consider the matrix elements in this formula. In the previous Appendix we calculated the eigenfunctions of the defect stripe and of the perfect Hamiltonian. However, these expressions are not normalized in real space. We now supply the proper normalization and evaluate the matrix elements. 

For the band eigenfunctions, we have plane waves with normalization factor $1/\sqrt{L_x L_y}$. Correspondingly, the matrix element reads
\begin{gather}
\langle v|...|c\rangle\!=\!\int...\langle v|r\rangle\langle r|c\rangle d^2 r\!\propto\!\frac{\delta\left(k_x-k_x'\right)\delta\left(k_y-k_y'\right)}{L_x L_y},
\end{gather}
where the delta functions express the momentum conservation law for optical transitions. After switching to dimensionless units $k_x,k_y\to a_x k_x,a_y k_y$, the square of the matrix element acquires the factor $\left(a_x a_y/L_x L_y\right)^2$. 

The defect state is localized along the $x$-axis and delocalized along the $y$-axis. Normalization along $y$ gives a factor $1/\sqrt{L_y}$. For the $x$-direction, we assume a localized wavefunction of the form $\exp\left(-|x|/a\right)$, where $a$ is the localization length (with the defect placed at $x=0$ for simplicity). Integrating over $x$ yields the normalization factor $1/\sqrt{a}$. Hence the total normalization factor is $1/\sqrt{a L_x L_y}$. For the corresponding matrix element we have
\begin{gather}
\langle d|...|c\rangle=\int...\langle d|r\rangle\langle r|c\rangle d^2 r=\nonumber\\
=\frac{\delta\left(k_y-k_y'\right)}{\sqrt{a L_x}L_y}\int...\langle d|x\rangle\langle x|c\rangle dx,
\end{gather}
where the delta function ensures momentum conservation along the $y$-axis. Integration over $x$ gives
\begin{gather}
\int dx \exp\left(ik_xx-|x|/a\right)=\frac{2a}{1+k_x^2a^2}.
\end{gather}
After integrating over $k_y'$ and switching to dimensionless units $k_x,k_y\to a_x k_x,a_y k_y$, the square of the matrix element acquires the factor
\begin{gather}
\frac{a_y^2}{L_x L_y^2}\frac{4a}{\left(1+k_x^2a^2\right)^2}.
\end{gather}
This implies that only small values of $k_x$ give a significant contribution to transitions between defect and valence bands. In addition, we can estimate the ratio of the factors for transitions to the conduction band and to the defect band
\begin{gather}
\frac{\text{v}\to\text{c}}{\text{v}\to\text{d}}\propto\frac{a_x}{a}
\end{gather}

\bibliography{apssamp}
\end{document}